\documentclass[authoryear]{elsarticle}

\usepackage[margin=2.5cm]{geometry}
\usepackage{graphicx}
\usepackage{subcaption}
\usepackage{xcolor}
\usepackage{url}
\usepackage{amsmath, amsthm, amssymb}
\usepackage{booktabs}
\usepackage{array}

\makeatletter
\def\ps@pprintTitle{%
 \let\@oddhead\@empty
 \let\@evenhead\@empty
 \def\@oddfoot{}%
 \let\@evenfoot\@oddfoot}
\makeatother

\usepackage[normalem]{ulem} 

\newcommand{\correction}[1]{#1}

\newenvironment{correctionenv}{\color{black}}{\ignorespacesafterend}

\graphicspath{{figures/}}

\begin{document}

\title{A multi-layer model for long-term KPI alignment forecasts  for the air transportation system}

\author[westminster]{Luis Delgado\fnref{fn1}}
\author[westminster]{G\'{e}rald Gurtner\corref{correspondingauthor}\fnref{fn1}}
\author[westminster]{Andrew Cook}
\author[innaxis]{Jorge Mart\'{i}n}
\author[innaxis]{Samuel Crist\'{o}bal}

\address[westminster]{School of Architecture and Cities, University of Westminster, 35 Marylebone Road, London NW1 5LS, United Kingdom}
\address[innaxis]{The Innaxis Foundation and Research Institute, G\'{e}nova 11, 2D. 28004 Madrid, Spain}

\cortext[correspondingauthor]{Corresponding author}
\fntext[fn1]{These authors have contributed equally to the work.}

\begin{abstract}

\correction{
This article presents a new, holistic model for the air traffic management system, built during the Vista project. The model is an agent-driven simulator, featuring various stakeholders such as the Network Manager and airlines. It is a microscopic model based on individual passenger itineraries in Europe during one day of operations. The article focuses on the technical description of the model, including data and calibration issues, and presents selected key results for 2035 and 2050. In particular, we show clear trends regarding emissions, delay reduction, uncertainty, and increasing airline schedule buffers.}

\end{abstract}

\maketitle

\section{Introduction and objectives}

The air transportation system is a complex system that is increasingly performance driven. Following the work of ICAO \citep{ICAO_performance}, SESAR has adopted and developed high-level KPIs (key performance indicators) in order to monitor its evolution and drive it in the desired direction. 

One of the issues encountered is the lack of ability to forecast trade-offs or alignments between KPIs. Whilst it is clear that these KPIs are not independent, it is less clear to what extent. A difficulty often facing KPI definition is finding sufficiently simple metrics that are readily understandable.

The reason why KPIs are so intertwined is that air transportation is a complex socio-economic system with multiple levels of dynamical processes, led by multiple actors with idiosyncratic rules and behaviours. As a consequence, a change in the system can have far-reaching consequences. Several changes at the same time can be interrelated and produce unforeseen impacts. This is a typical instance of an emergent system, as noted multiple times in recent years \citep{Bouarfa, Blom}.

This implies that partial models might not capture overall system performance, partially because the sum of their behaviour is not the behaviour of their sum. As a consequence, holistic models are required, which usually trade a high level of detail against a better integration of several mechanisms. The Vista project has built such a model to address this issue. The model simulates typical days of operation, representing different futures, at the current (2014), 2035 and 2050 horizons.

The project is mainly interested in the forecast of `what-if' scenarios for these horizons, i.e., we are aiming at taking into account various key factors and modelling how they shape the future. These factors, even though they are numerous, do not represent the full extent of the changes that will actually occur in the future. In particular, we are interested in (quite) incremental changes of the system, and to see what would be the state of the system with these changes. We purposefully do not try to predict what would happen with foreseeable major disruptions (such as drone operations), and obviously cannot predict the impact of unforeseeable ones.

The first period of Vista was dedicated to defining the scope of the model. In particular, the forces behind the potential changes in the system were identified (and called `factors' in the following), using an extensive literature review. Scenarios were also defined to prioritise the simulations and obtain knowledge on the most important factors. This has been described in \cite{sid_paper_2017} and is only briefly recalled here. In this article, we present an overview of the full model, the methods and data used for calibration, and some examples of the results obtained.

The paper is organised as follows. Section \ref{sec:operational-scope} explains the scope of the model, the main challenges it aims to address, and the scenarios chosen to be run. Section \ref{sec:description} describes the different models, while Section \ref{sec:calibration} explains how they have been calibrated. Section \ref{sec:results} presents some selected results obtained with the model from different scenarios. Finally, we draw some conclusions in Section \ref{sec:conclusions}.

\section{Operational scope}
\label{sec:operational-scope}

The Vista project is looking far in advance by trying to predict how the envisioned changes for 2035 and 2050 will likely interact. Given the extreme depth of the time horizon, Vista's aim is not to forecast exactly what will happen. Instead, Vista built a model that simulates a typical day of operations given a set of external forces (i.e., parameters that affect the system) such as macroeconomic, technological or regulatory parameters. By analysing the different indicators under different factors, Vista is able to identify trade-offs and trends across different timeframes and across different stakeholders for a given time horizon.

Five stakeholders are modelled in Vista: ANSPs, airports, airlines, passengers, and the environment. Vista considers KPIs for each of them in order to analyse the trade-offs when different factors are considered.

ANSPs are heavily regulated and have traditionally provided the full scope of air navigation services. The evolution to more performance-driven operations leads to a tendency towards the unbundling of services and technological innovation. In addition, almost all ANSPs have become engaged in one or more strategic alliances and industrial partnerships \citep{helios}. 
ANSPs are modelled in Vista as non-profit driven and, therefore, the capacity provided is based on the expected delay. The operating costs, and hence unit rates, are an outcome of these capacities and demands.

Large airports' current business models rely heavily on non-aeronautical revenues (parking, shopping, etc.) \citep{alfonso}. Congestion is a major issue for most, and different strategies are implemented to increase their capacity, such as soft management procedures or heavy changes in infrastructure \citep{berster}, or improvements from airport expansion programmes and technological enhancements. 
For small airports, aeronautical revenues are more significant and low-cost, point-to-point operations are more relevant to their income strategies. The relationship between airport operations and airlines business models results in the fact that the evolution of airports relies heavily on the business models of airlines and their future traffic. A spectrum of private and public ownership exists, but nearly all are heavily regulated, in particular regarding aeronautical charges \citep{adler}. Vista is able to capture both economic (e.g., revenues) and performance (e.g., delay) KPIs for airports.

Competition among airlines is an important factor when considering future traffic evolution. Airlines are highly market-driven as it is relatively easy to reassign aircraft to more suitable routes according to their business needs adjusting to changes in demand. Low-cost carriers (LCCs) generally have lower yields compared with `legacy' operators. LCC expansion has mainly been based on point-to-point strategies, aiming at higher utilisation by using a homogeneous fleet, and lower costs by using secondary airports \citep{Doganis}. However, more recently, some LCCs have shifted to the legacy model to some extent by operating from primary airports or feeding long-haul flights. 
Legacy carriers have been forced to adapt their model by lowering their costs, sometimes trying to gain market share with an `in-house LCC', and by unbundling services provided following an LCC approach to pricing. Vista can capture different indicators for this stakeholder at different ATM planning phases.

For the passenger, price, travel time, comfort and convenience constitute some of the factors influencing their choice \citep{dataset_d3_2}. The literature often defines archetypal profiles for passengers, usually taking into account socio-economics and travel purpose (often simply `business' or `leisure'). Some of these profiles have been defined by the project DATASET2050 \citep{kluge} and have been loosely adopted in the Vista model. This more detailed passenger profiling allows us to model the door-to-gate and gate-to-door phases providing door-to-door metrics, in addition to more classical gate-to-gate indicators.

The final `stakeholder' modelled in Vista is the environment. This is a passive stakeholder, which interest lies on the reduction of the impact of aviation on the environment. In particular, metrics related to the emissions of (CO$_{2}$ and NO$_{x}$). At this stage, noise is not included in Vista.

The selection of scenarios run is based on a consultation, a dedicated, expert workshop and the current model's capabilities. The way scenarios are used in Vista has been explained in \citep{sid_paper_2017}. Here, only the main features are highlighted.

Scenarios are built by fixing exogenous variables to the model. In Vista, these variables are called business or regulatory `factors', and include macro-economic considerations such as GDP, but also SESAR solutions such as free routing. These factors are gathered into sets and fixed at certain values for a given scenario.

Nine scenarios, which are presented in Table~\ref{tab:scenarios}, have been modelled in Vista. The first scenario represents the `current' situation (2014), used for calibration and comparison. We then used two main baselines: the `Low' baseline is built around slow economic growth and slow technological improvement. The `High' baseline, in contrast, includes high economic growth and technological improvement.

\begin{table}[htbp]
\begin{center}
\caption{Examples of scenarios simulated in Vista.}
\label{tab:scenarios}
\begin{tabular}{c|c}
\toprule
Scenario & Short description\\
\hline
\hline
Current & `Current' situation (SEP 2014) \\
\hline 
L35 baseline & 
\begin{tabular}{c}
Baseline environment in 2035 \\ (slow economic growth and\\ slow technological advancements)
\end{tabular}\\
\hline 
H35 baseline & 
\begin{tabular}{c}
Baseline environment in 2035 \\ (high economic growth and\\ high technological advancements)
\end{tabular}\\
\hline
Non-supportive 2035 & 
\begin{tabular}{c}
Using L35 baseline plus a poor emphasis on \\environmental and passenger protection \\and very a high price for fuel
\end{tabular}\\
\hline
Supportive 2035 & 
\begin{tabular}{c}
Using L35 baseline plus a poor emphasis on \\environmental and passenger protection \\and very a high price for fuel
\end{tabular}\\
\hline
\begin{tabular}{c}
L50, H50, \\
Non-supportive 2050,\\
 Supportive 2050
\end{tabular}
& As per above, for 2050\\
\hline
\end{tabular}
\end{center}
\end{table}

In addition to the baseline scenarios, four factors have been selected in order to assess their impact on the system. These factors are grouped into `supportive' and `non-supportive' cases. Loosely speaking, the former depicts a world where passenger rights are protected, environmental issues are tackled and airlines benefit from a favourable economic situation (low price of fuel). The latter describes the opposite situation. In brief, the factors are set as follows:
\begin{itemize}
\item price of fuel: low in supportive; high in non-supportive;
\item implementation of passenger reaccomodation tools and provision schemes: `on' in supportive; `off' in non-supportive;
\item implementation of 4D trajectories ECAC-wide: current operations in non-supportive; fully implemented in supportive;
\item price of emissions allowances: high in supportive; low in non-supportive.
\end{itemize}
NO$_X$ emissions are also taxed (based on their equivalent radiative impact compared to CO$_2$) in the supportive case, and are not taxed in the non-supportive case. Note that in this article we are not considering the possibility of the introduction of biofuel, which would change the net emission of the flights. It is unlikely that biofuels will be introduced at a large scale in the next decade, although this may be the case by 2035 or 2050 \citep{KOUSOULIDOU2016166}.

The way in which the parameters defining the scenarios evolve, varies across the different types thereof. Macroscopic parameters, such as GDP, are taken from a cross-section of forecasts. Free parameters, such as the price of fuel, are set to what is expected to be `high' or `low' or values, based on market fluctuations and trends.

A third type of parameter is set by using SESAR high-level plans. Indeed, the SJU has defined over the years different key performance indicators (KPIs), with associated targets. These targets used to be set based on the ``time-'', ``trajectory-'' and ``performance-based'' operational concept \citep{MP-1}. SESAR has since moved away from this nomenclature \citep{MP-2}, but we used these targets to change the parameters, assuming that they represented a best estimate for the future, albeit maybe too optimistic. We collected the targets, and achieved performance, for each former SESAR subpackage, mapped them to exogenous variables in the model, and modified the latter based on the changes detected in the former. Different assumptions were used for each scenario, for instance considering that trajectory-based operations are reached only by 2050 in the low baseline, but that performance-based operations are implemented in the high baseline by 2050. Moreover, we performed extrapolation for some KPIs based on the evolution of other indicators, where data were missing. As a concrete example of how we used these numbers, we considered the 9\% (summed over different subpackages) target in the airport capacity indicator planned for time-based operations to increase the capacity of airports in the model by 9\% in 2035 in the low scenario.

\section{Description of the model}
\label{sec:description}

\begin{figure}[htbp]
\begin{center}
\includegraphics[width=0.7\textwidth]{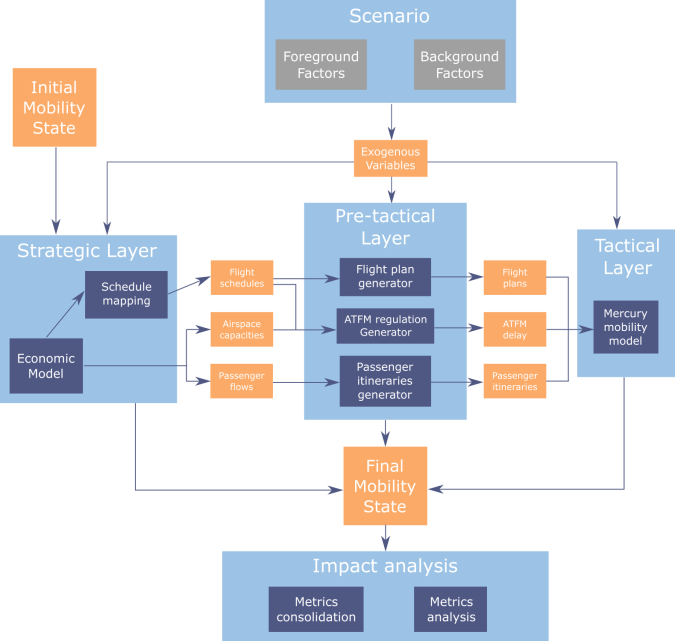}
\caption{Vista model schema.}
\label{fig:schema}
\end{center}
\end{figure}

\correction{Figure~\ref{fig:schema} presents an overview of the Vista model. Vista models the three temporal phases of ATM (strategic, pre-tactical and tactical) for each scenario investigated, with the objective of generating a representative (busy) day of operations for each given scenario. The various factors define the scenario to be modelled. The strategic layer considers the factors and the economic environment to provide the outcome of strategic decisions made by the stakeholders, the capacities provided, demand, flight schedules and passenger flows for a typical day of operations. These flows, schedules and capacities are transformed into individual flight plans, passenger itineraries and ATFM regulations by the pre-tactical layer. Finally, the tactical layer executes the flights and passenger itineraries at a flight and passenger level, tracking the evolution of delay, passenger connections and the tactical decisions carried out by the actors. Among these layers, only the tactical one is based on a prior model, `Mercury'. Mercury has been re-implemented and enhanced for Vista, but it is based on previous models, used during the last ten years across several research projects such as POEM and ComplexityCosts~\citep{ComplexityCost}. A new version of Mercury, based on a powerful agent-based paradigm, is currently being developed in the ER3 project Domino \citep{DominoSID2018, DominoSID2019, MAZZARISI2020101801}, to which the reader is referred for further developmental details.}


One characteristic of Vista is that each layer may be executed integrated into the full model, or as a stand-alone model, which allows us to test the different phases independently, by providing the appropriate inputs.

\subsection{Strategic}
\subsubsection{Economic model} 

The economic model is the first block of the strategic layer and has the task of creating appropriate levels of supply and demand in Europe based on different scenarios. The model provides flows of flights and passengers for each origin-destination pair.





The model is based on a common environment, in which different agents evolve. The types of agent implemented in the model are summarised below. 
\begin{itemize}
\item Airline: with one instance per actual airline, the airline is used to compute and handle flights.
\item Flight: with one instance per OD pair (without connection) per airline, the flight object notionally represents all the flights operated by an airline between the OD pair. It computes the marginal profit of the leg and chooses the supply (number of seats) for the next simulation turn. It is also handles choices between the possible flight plans based on their fuel and ATC charges costs. 
\item ANSP: with one instance per actual ANSP, ANSPs set their capacities (number of controllers) based on a target delay. They then set the unit rates to have zero net profit (see below) based on the expected demand. 
\item Passenger: with one instance per initial origin `O' and final destination `D' (with connections), the passenger agent notionally represents all the passengers going from O to D by any legitimate itinerary (see below). 
\end{itemize}
There are no hard-coded archetypes of agents within each type. Instead, the different behaviours of the agents are defined by their cost structure and their initial conditions (the initial network for airlines, for instance). As noted above, a (dynamic) network underlies the structure of the model, where the airports are nodes and flights are edges. In this network, passengers use collections of edges from their origin to their final destination.

Supply and demand interact in this network in an intricate way. On the one hand, the supply is leg-based, each airline creating its own capacity for each leg. To compare the supply and demand, supply is then aggregated based on all the passengers going through this leg for this airline, as part of any `itinerary' (an `itinerary' is a distinct permutation of legs between a given OD pair, operated by a given alliance). On the other hand, demand reacts to the prices of itineraries. These prices are thus aggregated for each leg (using a simple summation rule) in the corresponding itinerary. 

At each turn, the agents perform a number of actions depending on their nature, see \citep{del_52} and below for more details. In brief: 
\begin{itemize}
\item based on the scenario, the parameters of the agents are changed; 
\item ANSPs predict the traffic for the round; they compute the capacity needed to be under a target delay per flight and set this value for their capacity, limited by their maximum capacity; they set their unit rate so that their expected profit is null; 
\item flights predict delay at airports, en-route (ATFM) delay, and the price of their leg for this turn; they set their supply based on these predictions and their own cost function; 
\item passengers take the price of the last round for each leg and sum them to form the price for the itineraries; they weight each itinerary based on their utility function (see below) and set their demand level for each of them based on this weighting; 
\item demand is aggregated for each leg; 
\item supply and demand are compared for each leg - prices evolve on each leg based on the discrepancy; 
\item all agents record variable values for this round, i.e., prices, delays, etc; 
\item a new turn is initiated.
\end{itemize}

The way in which agents make their decisions is influenced in particular by the technological developments set in the scenarios.

For example, the ANSP is a profit-neutral entity that adjusts its capacity in order to achieve the set delay targets, setting the corresponding unit rate to cover its cost. Indeed, the ANSPs are similar to airports, in the sense that they `produce' some delay for flights, based on their level of traffic. The relationship between the increase of delay and an increase of traffic implicitly defines their capacity $C$. ANSPs are, under current SES legislation, supposed to keep the average delay per flight under a certain threshold, that in the following we call the `target delay', which should be understood as the `target maximum delay', with an associated cost efficiency (driven downwards in the long term by the legislation).

Thus, we assume that, given a fixed structure of delay and traffic forecast (for the next iteration), the ANSP is able to adjust its capacity so as to reduce the average delay below the target. Given its cost efficiency target, the ANSP is then able to compute the unit rate it needs to cover its cost, based on the traffic prediction. Note that both the maximum capacity and the cost efficiency depend on external factors. In Annex~\ref{annex:ANSP}, we give more details on the process, including the assumptions regarding the structure of delay and how the parameters are calibrated. 

It is important to note that we are not suggesting that ANSPs change their capacity on a daily basis. We use a single (representative, busy, undisrupted) day, mainly due to data limitations, that we assume is representative for a longer period of operation, i.e., a year. The ANSPs then take decisions on their capacity based on the evolution of the metrics recorded on that representative day. In particular, we do not assume that there is a mappable relationship between a single iteration and a given period of time. We are only interested in the final state of the model: the long-term equilibrium.

There are various limitations arising from these assumptions. An important one is that we assume that ANSPs have enough flexibility to change their capacities at the two horizons of interest, 2035 and 2050. This might not be true, for instance because of various frictions (technology uptake, labour management, etc). Another limitation is the fact that the representative (busy) day may be less representative in the future, because schedule characteristics and seasonalities, for example, change throughout the years. Further, some ANSPs in reality are known to delay investments for capacity improvements, and thus generate a surplus (whilst others make capital overspends). This is not reflected in the model, but it can be considered as another type of friction. The frictions are neglected based on the fact that we are interested in long-term horizons, for which they hopefully play a lesser role, driven by longer-term regulatory review, controlling (rolling) surpluses and driving better cost efficiencies.


The airport uses the same kind of delay-capacity relationship, but only passively. Indeed, since capacity extension for an airport is a very complex issue, and highly context-dependent, we assumed that the capacities were constant, save for the technological improvements. For given average traffic, the airport thus creates an average delay, based on a relationship calibrated on empirical data. Following the literature, we use a linear law:
$$
\bar{\delta t} = \delta t_0 + \frac{T}{C},
$$
where $\bar{\delta t}$ is the average delay `created' at the airport, $\delta t_0$ and $C$ are constants, with $C$ representing the capacity of the airport in a given time window, to be compared with the (mixed) traffic $T$ during this time. This form and the data used to perform the regression in order to obtain the values of $\delta t_0$ and $C$ (for each airport) are further described in Annex~\ref{annex:airport}. Any increase in capacity, driven by the SESAR airport capacity indicator, is thus translated into a decrease  of the average delay for given average traffic.

Note that the strategic model is deterministic, i.e., agents take deterministic decisions and no random noise is added to the system. This choice was mainly driven by the computational requirements. All the subsequent models and submodels are stochastic, which means that many iterations have to be performed in order to have meaningful results. The fact that the first model is deterministic allows us to avoid performing a \textit{very} high volume of such iterations.

The airlines also use continuous variables for their supply, i.e., they have real numbers of seats. The demand is also continuous, which allows a smooth convergence of the model and a simpler algorithmic setup. The results are then discretised to be used in subsequent models.

\subsubsection{Schedule mapper}

The schedule mapper is the second block in the strategic layer. It converts the high-level flows of the economic model into individual schedules, to be used by
the flight plan generator. 
Planning schedules based on expected demand is a highly demanding task, even at the single airline level. Airlines usually have dedicated tools for this. The complexity of assigning schedules is due to the high number of possibilities and the multiple constraints. These constraints include hard constraints such as crew, aircraft, and airport slots, plus soft constraints such as the cost of operating an OD pair. It is out of the scope of Vista to reassign completely from scratch all the schedules created by the economic model. In particular, this would imply capturing very complicated slot management behaviours from the airlines (including `irrational' behaviours such as endowment effects). Moreover, the number of flights considered is too large to ensure a sufficiently fast, stable solution from a computational perspective. Vista simplified the problem by relying on an initial state, set to be 12SEP14 (see later).

The schedule mapper compares, for each airline, the new flow to the old one for each OD pair. If the new flows are high enough compared to the old ones, each airline tries to add a new aircraft and optimise its route so that most of the new flow is covered. If the new flows are small, it removes one of its aircraft. The airline takes crew and airport slots into account only indirectly through the possible patterns (routes) available to the new aircraft, and the corresponding turnaround times (see later, under `Calibration'.).

More specifically, the mapper goes through the following steps:
\begin{itemize}
\item load data on airports, historical schedules, pattern data (see Section~\ref{sec:analysis}), and strategic flows;
\item compute average travelling times between every OD pair;
\item compute likely departure times;
\item load the decision tree for the turnaround times;
\item for each airline:
\begin{itemize}
\item trim its network by removing aircraft which are in excess;
\item grow the network by adding aircraft to meet demand;
\end{itemize}
\item compute the new schedules and add them to the database.
\end{itemize}

\subsection{Pre-tactical}

\begin{figure}[htbp]
\begin{center}
\includegraphics[width=0.7\textwidth]{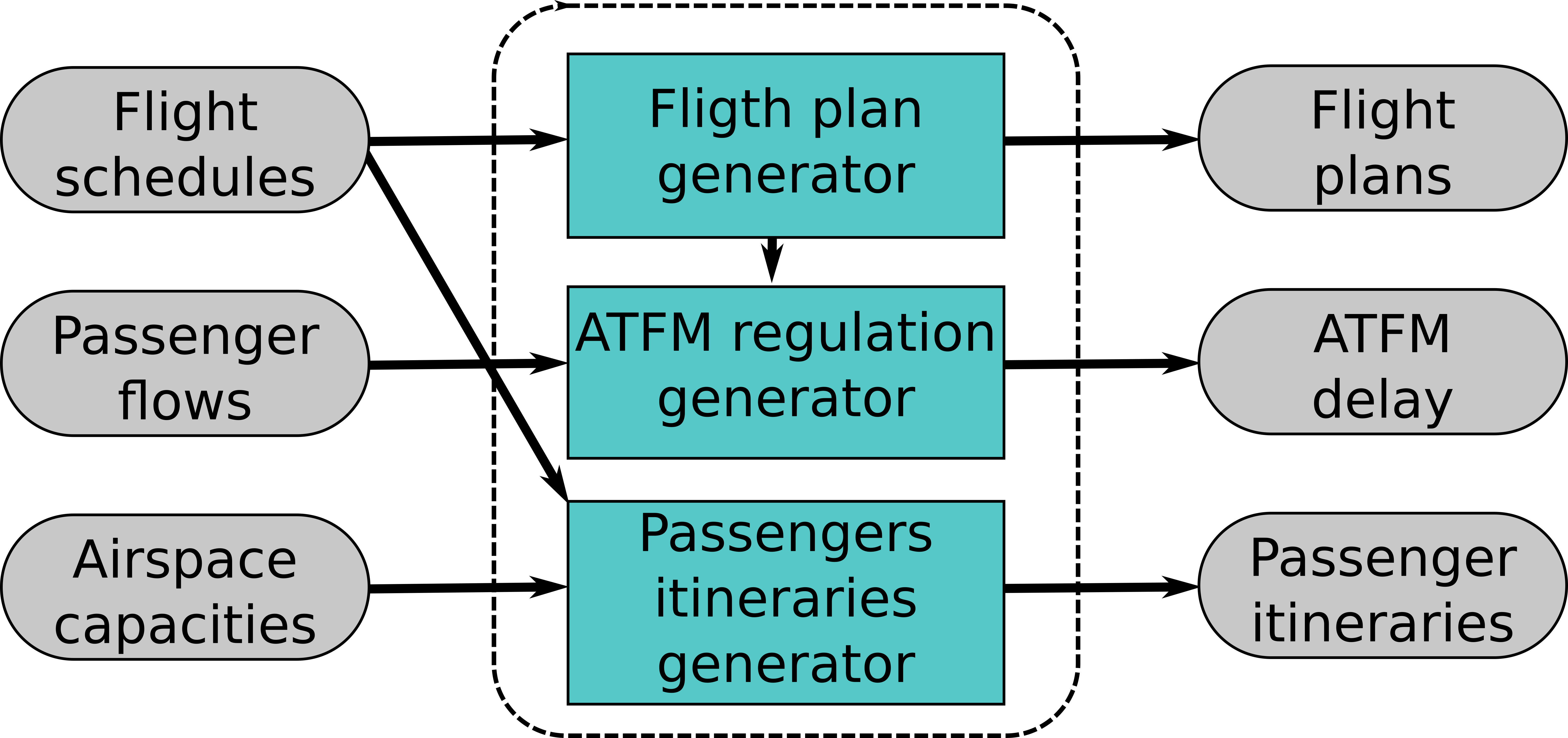}
\caption{High level pre-tactical layer.}
\label{fig:pre_tactical_high}
\end{center}
\end{figure}

As shown in Figure~\ref{fig:pre_tactical_high}, the objective of the pre-tactical layer is to create the necessary level of detail to execute tactically the flights and passengers' itineraries. The outcome of this layer are individual flight plans, ATFM regulations and probabilities of being assigned delay due to a regulation, and individual passengers' itineraries.\correction{ The layer is divided into three blocks each one of them devoted for each of the previously mentioned outputs:}
\begin{itemize}
    \item Flight plan generator;
    \item ATFM regulation generator;
    \item Passengers' itineraries generator.
\end{itemize}

\subsubsection{Flight plan generator}

The flight plan generator transforms origin-destination schedules into possible flight plans. For each flight plan, an estimated operating cost is computed including fuel, en-route airspace charges and emissions costs. These operating costs are considered when prioritising the different flight plan options for each schedule. This block is based on historical (possible) trajectories and on aircraft performances.

The flight plan generation is divided into different sub-processes 
supported by historical data analysis on flight trajectories and aircraft performance and structured in different processing blocks:

\begin{itemize}
\item Route generator: for each scheduled OD pair, a set of possible routes is computed. These routes are based on the clustering of historical routes (see Section~\ref{sec:analysis} for more details on this clustering). If the OD pair does not exist in the historical dataset, new routes are estimated. This is done considering existing routes between the ANSP of the origin and destination airports. The routes which are closer to the OD under study are selected and extended as required. If no flights exist in the historical dataset linking the origin and destination ANSP, the possibility of estimating the route considering an intermediate airport is analysed. Great circle distances between entry and exit points of the ANSP are increased considering an inefficiency factor estimated from historical data to account for airspace inefficiencies. Note that factors from the scenarios that might affect the length of the route are considered. For example, in the case regarding the implementation of free routes, this is translated into a reduction of the total flight plan length.

\item Trajectory generator: the next step is for each possible route to compute a trajectory. This trajectory includes, \emph{inter alia}, climb, cruise and descent distances, selected altitudes, air speeds, estimated fuel usage and average cruise wind. I.e., we extend the two dimensional route into a full, four dimensional trajectory. A trajectory optimiser could have been used to generate the different flight plans. However, with that approach idealised optimal trajectories would have been obtained. Instead, in Vista, the trajectory generator relies on historical data on flight plans and aircraft performance (BADA), obtaining \emph{realistic} trajectories. 
The different processes required to generate these trajectories are: the estimation of average cruise winds, the selection of flight level and cruise speed, and the definition of the climb and descent profiles. Some of the processes are stochastic, therefore, this trajectory generator block could be executed several times in order to obtain different possible trajectories per schedule and route:
\begin{itemize}
    \item Average cruise flight level: an average flight level for cruising is assigned to each route based on historical distributions of flight level requested per flight, as a function of aircraft type and flight plan length.
    \item Climb and descent phases: the climb and descent distances are estimated based on information on aircraft type and flight plan distance. With this, the cruise phase length can be computed. The required time for the climb and descent are estimated using fittings between these phases' distances and times from historical data.
    \item Cruise phase: the average cruise speed request is estimated considering a historically estimated distribution of speeds requested as a function of cruising flight level. This speed request is done ensuring that the value is within the aircraft's limitations. For each origin and destination ANSP an average cruise wind distribution is estimated, from the analysis of historical datasets, and based on these distributions, the average cruise wind is assigned for each trajectory.
    \item Fuel estimation: using BADA performance data, the required climb and descent fuel are computed. Considering these values along with the cruise distance, flight level, cruising speed, wind and an estimation of the landing weight, based on expected payload, the fuel and the average cruise weight are estimated. A conversion between fuel burn and $CO_{2}$ are also estimated per trajectory so that the expected cost of emissions can be considered.
\end{itemize}

\item Flight plan generator: computes, for each trajectory, its estimated direct operating costs: en-route airspace charges, cost of fuel and emissions. The en-route airspace charges are computed modelling the 39 regions managed by EUROCONTROL CRCO plus the airspaces of Egypt, Belarus, Morocco, Uzbekistan and Ukraine. Other surrounding countries which follow different charging schemes are also modelled: Algeria, Iceland, Russia, Tunisia. This allows us to compare the cost of different routes even when they use adjacent airspaces to the core European ones (for example flying over Algeria to cover a route between the Canary Islands and Italy or, using the Shanwick airspace in flights between the UK and the Canary Islands). The cost of the unit rate used for the core ANSPs considered in Vista are either from historical values or the outcome of the economic model of the strategic layer. The values of fuel and carbon emissions also depends on the scenario considered.

\item Flight plan selector: the trajectory generator produces as many trajectories as routes between the OD pairs. However, the tactical layer requires a single flight plan per schedule. This flight plan per schedule is also required by the ATFM regulation generator to estimate the demand on each ANSP and, from this, the probability of having an ATFM regulation. However, during the tactical operation of the flight, airspace users often change their flight plan prior to departure as a function of tactical conditions. These tactical changes to the flight plans cannot be decided in the pre-tactical layer. The flight plan selector prioritises all the flight plan options of the flights based on their operating cost. A logit rule, which considers the cost of fuel, of airspace charges and of emissions is used to estimate the probability of selecting a specific flight plan, in a similar manner to the airline computing the cost of the flight plan in the strategic layer.
\end{itemize}

All the blocks of the flight plan generator might be affected by the various factors of the environment. For example, the length of the routes used by the route generator will be affected by the introduction of free routes, or the estimation of the cost will depend on parameters such as fuel costs or route charges, which would have been previously defined by the strategic layer.

\subsubsection{ATFM regulation generator}
\label{subsec:ATFM_generator}

The ATFM regulation generator estimates the probability of being affected by ATFM regulations and the corresponding amount of delay. As presented in Figure~\ref{fig:pre_tactical_high}, the input of the ATFM regulation generator is the capacity of the ANSPs (for the scenario to be processed, and for the baseline 2014 case) and the traffic (estimated demand of the scenario to be processed, and of the baseline 2014 case). The ATFM generator thus requires the outcome of the flight plan generator (and flight plan selector) and the 2014 demand and capacity, in order to be able to compute the variation of demand and capacity with respect to the 2014 baseline case, which is used as a reference.

ATFM regulations are divided between regulations due to capacity issues (i.e., regulations flagged as ``C'' when issued by an FMP) and all the other regulations. It is assumed that regulations due to capacity are affected by the demand on the ANSP, while the regulations that are not due to capacity are independent of the demand and related to other operational factors, and hence maintain the same probability of being issued across the ECAC region as in the historical data. This assumption allows us to modify the probability of having ATFM delay due to the demand expected at different ANSPs and their expected capacity, while maintaining delay, which is not directly linked with capacity/demand imbalances (e.g., regulations due to weather). From historical data (analysis of AIRAC 1313-1413), regulations due to capacity represent 43.3\% of all the regulations issued, followed by weather (25.8\%) as the main causes. Weather, by its nature, might occur at wider locations across the ECAC and might not be related directly to demand.

An experimental, cumulative probability distribution function, based on the analysis of a year of historical regulations (AIRACs 1313-1413), is used to assign the amount of delay. 

If a flight is not affected by a regulation issued for non-capacity demand imbalance issues, then there is a probability of incurring a delay due to an ATFM regulation due to capacity, which depends on the ANSP airspace that the flight crosses. Each ANSP has a given probability of assigning delay, which depends on the time the flight is crossing the ANSP’s airspace and on the expected demand and capacity. 
The probabilities of delay due to capacity are based on historical analyses of flights affected by regulations in each ANSP. These probabilities are adjusted considering the variation on demand and capacity offered by the ANSPs, considering the factors modelled in the different scenarios as described by
$$
P = P_0\left(\frac{(D_1-D_0)}{D_0}-\frac{(C_1-C_0)}{C_0}\right)+P_0    
$$
where $D_0$ and $C_0$ are the demand and capacities from the baseline 2014 for the ANSP, $P_0$ is the probability of delay when crossing the ANSP’s airspace due to capacity issues in the 2014 baseline scenario, and $D_1$ and $C_1$ is the demand and capacity in the scenario modelled. The capacities ($C_0$ and $C_1$) are part of the outcome of the economic model in the strategic layer, and the demand is estimated by counting entry counts in 30-minute windows.

\subsubsection{Passengers itineraries generator}
\label{sec:pax_generator}

The strategic layer generates the passenger flows and flight schedules (including their aircraft type and number of seats available). The objective of the pre-tactical layer is to transform these passengers' flows into individual itineraries, i.e., to assign the passengers to specific flights, while preserving aircraft capacity and calibrating with historical datasets such as load factors, as described in Section~\ref{sec:pax_calibration}. This is done in a three-stage process:
\begin{itemize}
    \item computing the possible options available for the passengers in each flow considering the minimum connecting times at the airports. In some cases, the same itineraries with the same airlines can be achieved with different flights, which produce different connecting times at the airports.
    \item optimising the assignment of passengers among their options considering aircraft capacities and minimum connecting times at airports. This has been achieved by formulating an integer programming problem where the objective function is to maximise the number of passengers that are assigned, while maintaining the number of passengers on the flights lower than the capacity of the flight, and not assigning more passengers from an itinerary flow than the volume of that given flow. The number of passengers to be assigned per flight is considered by assigning a target load factor per flight, which is set probabilistically considering calibration parameters. Note that this optimisation carries the risk that the optimiser might prioritise single-leg itineraries, as they are easier to accommodate, and to maximise the number of passengers allocated. However, we want to preserve as much as possible the two- and three-leg itineraries, as these are more relevant in terms of delay propagation through the network, and single-leg itineraries can be added afterwards as ‘fillers’ if needed (see below). For this reason, a two-stage optimisation process is implemented: first, an assignment of only connecting passengers (with two or three legs), then using the outcomes of this optimisation as constraints for the optimisation of all the flows.
    \item creating additional passengers' itineraries to ensure that the load factors of the aircraft are realistic. A new target, load factor occupancy is set on the flights as the maximum between the current load factor and a triangular distribution between 0.35, 0.8 and 0.8. This might mean that more itineraries are needed. These are then generated as ‘fillers’ of single-leg passengers on those flights. The number of passengers added as fillers is defined by the load factor of the aircraft. This means that there might be an expected discrepancy with respect to the number of passengers planned at the strategic level, even if this is small, as the capacity provided strategically is made considering the expected demand. (The number of filler passengers ranges between 2.5\% and 7.1\% of the total number of passengers in each scenario, being on average 5.7\% of the passengers in a given scenario). These filler passengers do not have a fare associated to them, therefore a linear fitting is used to estimate the cost of a given fare based on the fare of passengers doing single-leg itineraries and the flight distance. Some randomness is added to this value.
    \item assigning which passengers are `premium' and which are `standard' (which features in passenger reaccommodation rules in the tactical layer). Based on the fitting between fare and trip distance, passengers are classified between categories, with a probability depending on the extra price paid with respect to an average passenger on similar itineraries. The total number of `premium' passengers is adjusted to meet the calibration target.
\end{itemize}

\subsection{Tactical layer}
\label{sec:tact}

The tactical layer models delay propagation between flights and the adaptability of the system during disruption (cancellations, background and foreground delay) and with limited resources (e.g., airports and en-route capacity). The modelling is performed both per flight and per passenger to allow us to capture both flight and passenger indicators. The model used for the tactical layer is the Mercury event-driven simulator, as introduced earlier. One of the characteristics of this tactical model is its door-to-door capabilities. Passenger types are disaggregated into more detailed profiles (e.g., `environmental traveller', `culture seeker') and mapped to different transport choices regarding access/egress to/from the OD airports.

There are various feedback loops considered in this layer. One example is the expected arrival time of flights, which is updated to the airline several times during the simulation. The airline operator can use this information to adjust the behaviour of outgoing connecting flights, for example, by waiting for connecting passengers. 

Several airline costs are considered during the simulation: fuel, emissions, CRCO, crew, maintenance and passenger delay costs. Those have an impact on the behaviour of the system such as when selecting the flight plan. This allows us to compare how strategic planning by ANSPs and airlines evolves to specific flights, capacity-demand regulations and ANSP revenues in the pre-tactical layer, and, during the tactical phase, materialise as actual delays and costs. The model also updates the cost metrics considering the last state of the system (e.g., passenger compensation costs). Environmental emission costs are also considered, in particular $CO_{2}$ and $NO_{X}$ emission costs which are already considered in the strategic layer.

The tactical layer simulation is a sequence of two processes:
\begin{itemize}
    \item the gate-to-gate simulation, modelled on an event-driven simulator which considers information from previous layers, such as the flight schedules, flight plan options, ATFM delay and passenger itineraries, along with other data such as expected connecting times for passengers, taxi times, or turnaround processes times for different aircraft types at different airports. There are a total of four events modelled: 
    \begin{itemize}
        \item flight plan submission: which models the tasks related to flight plan selection and submission, including processes such as cancellation of flights, ATFM delay assignment with the modelling of Network Manager activities, which considers the probabilities of assigning ATFM delay as estimated in the pre-tactical layer, and the estimation of non-ATFM delay. The flight plan contains the route and the ANSPs that will be crossed, with their estimated entry times. If the flight departs from an airport located in an ANSP that can assign delay due to ATFM (i.e., ECAC, EUROCONTROL and adjacent states with regulated traffic incoming to the NM area as defined in \citep{ECTL_assumptions}) then it can potentially incur ATFM delay. This delay will be first drawn from the probability of having ATFM delay not due to capacity issues. If this is not the case, the probability of having delay assigned per ANSP crossed, as estimated in the pre-tactical layer, will be considered.
        \item previous aircraft ready: once the turnaround process is finished, this event is triggered to explicitly compute metrics such as reactionary delay. If an ATFM slot has been assigned to the subsequent flight, this might trigger a request for a new slot if the previous one is missed.
        \item push-back ready: at this point, the processes related to the trajectory execution are computed. This event also triggers the reaccommodation of passengers who missed a connection.
        \item arrival processes: once the flight arrives at the TMA of the arrival airport, this event manages the arrival sequencing (e.g., assigning holding delay if required) and the taxi-in processes.
    \end{itemize}
    Aircraft are assigned to flights and therefore reactionary delay is explicitly considered in the model.
    
    \item the door-to-door simulation (which is computed as a post-processing analysis adding estimated access/egress times from the airports and required times at the airports for processes relating to the different simulated gate-to-gate passenger itineraries), is as described in \citep{dataset_d5_2}.
\end{itemize}

\section{Calibration}
\label{sec:calibration}

The full calibration of the model required substantial data acquisition and analysis. In the following, we highlight the main sources and the main steps carried out. For current operations, the model is calibrated based on a day of traffic (12SEP14) selected due to being a busy, nominal day (e.g., not disrupted by industrial actions or severe weather events).

\subsection{Input data}
\label{subsec:calibration_input_data}

One of the main characteristics of the Vista project is the of use different sources of data to inform the model. The main data source is EUROCONTROL's DDR data~\citep{EUROCONTROL_DDR2}. The data were used extensively, in particular to:
\begin{itemize}
\item set the initial state of the economic model;
\item extract the distribution of delays for airports and ANSPs, which helped to:
	\begin{itemize}
	\item infer delay-traffic relationships (for airports);
	\item perform a mean-variance analysis on airport delay;
	\item perform an analysis of ATFM regulations;
	\end{itemize}
\item compute the length of trajectories in each ANSP's area;
\item cluster possible routes between origin and destination airports;
\item model flight plan preferences (flight level and speed requests);
\item model flight trajectories (characteristics of climb and descent phases);
\item estimate average wind distributions between regions by comparing ground speed with requested air speed.
\end{itemize} 
Other flight-related data have been used, in particular the BADA 4.2 model in order to estimate fuel consumption, both in the planning (pre-tactical) and executive (tactical) phase.

Since DDR data is flight-centric, we used other sources of data to compute passengers' information. We used a mixed database from global distribution system and IATA (`PaxIS') data to obtain detailed information on passengers' itineraries for 12SEP14, including the fare price and the class of the passenger. 
Passengers elasticities have been sourced from the literature (\cite{iata} for price and income, and \citep{Adler_2001} for frequency). To estimate the increase of passenger income and thus the increase in demand, we used projections on future GDP from \cite{Eurocontrol2013}. These projections have been extrapolated for some countries, using decreasing growth over time.

In order to build the available itineraries, Vista also used data relating to airline alliances and partnerships. To simplify the model, we considered that the flights of any airline in a partnership (or alliance) can be used in combination with any other flight from the partnership or airline to create a valid itinerary.

To estimate the different operational costs of the airlines, we used the computations of \cite{cost_delay}. This includes both strategic (planning, buffer) and tactical (delay) costs, and is used as a standard reference, e.g., by the Performance Review Body to estimate the cost of delay. 

Some financial data have also been used for airports. In particular, we used these data to estimate their costs and the landing and departure fees for the airlines.

For the ANSPs, we used only the assumption that they were supposed to be at zero profit and that they have to be under a target value for their delay per flight. Using the initial delays, we are able to compute some efficiency metrics for them (cost per unit of capacity). This efficiency is then changed based on the different scenarios considered.

Other data sources were required in order to model the distribution parameters for the different flight and passenger phases such as taxi times (EUROCONTROL (CFMU) datasets), minimum turnaround times, minimum connecting times and non-ATFM delays \citep{coda}. Door-to-gate and gate-to-door modelling relies on distributions calibrated in the DATASET2050 project \citep{dataset_d5_2}.


\subsection{Calibration analyses}\label{sec:analysis}

Various analyses were performed to calibrate the model. The most important are listed in this section.

\subsubsection{Pairs of variables analysis}
First, several linear and non-linear regressions have been performed to examine pairs of variables. For instance, the link between the mean delay at an airport and the level of traffic has been established using linear regression over different time windows and at different airports. The standard deviation of delay (i.e., the unpredictability in the model) has been computed using correlation analysis between the mean delay and its standard deviation. This analysis has produced some interesting results \emph {per se}, for example that mean delay is more correlated with traffic when the traffic is high at the airport.

\subsubsection{Route clustering}
In the pre-tactical layer, when airlines have to choose the horizontal trajectory (route) that their flights will follow, they use a reduced number of routes to choose from. This is done to limit the computation time of the algorithm, only sacrificing details in the model. Indeed, empirical data show that the flights plan lots of different routes in the airspace, but most of them are very similar, or identical. As a consequence, we use a clustering algorithm to extract the typical routes and use them in the model. 

In Vista, we are less focused on the low-level interaction of flights from take-off to landing (e.g., modelling conflicts and their resolutions). The flight phases (climb, cruise, descent) are computed based on the performance of the aircraft, route length and uncertainties added to the actualisation of the flight duration. These uncertainties capture the effects of these tactical interventions on the trajectory (e.g., conflict resolution, directs given en route). As explained in Section~\ref{subsec:ATFM_generator}, the probability of having ATFM delay assigned to a given flight is dependent on which ANSP airspace areas are crossed. The cost of operating a route will depend on the route charges, which vary as a function of these crossed areas. Between a given origin-destination pair, two routes can thus be considered `equivalent' in Vista if they have a `similar' length and cross the same airspace areas.

The clustering process starts with the set of historical flight routes. We group them based on which ANSPs are used and in which sequence (e.g., Spain-France-UK). This is to keep sufficient variability for a given sequence of ANSPs. Then, we used a clustering algorithm on each bundle to keep only a few routes per sequence.

Before clustering objects, one has to define a representation of them in one or more dimensions which allows to define a distance function. In this study, we use the total length of the route as the only attribute. The problem is thus uni-dimensional, which allows an efficient algorithm such as kernel density estimation (KDE) to be used. This simple setup is surprisingly good, and allows to extract typical, well-differentiated clusters in most cases. 


Using these typical trajectories, the algorithm then chooses the flight level, flight speed, and climb and descent times to produce the flight plan alternatives. This is done by classifying the different possible values for these parameters per type of aircraft and flight plan distance. The result is that individual flights have highly specific trajectories, but still compliant with the typical features observed in the historical data. Once the flight parameters have been selected from the historical data, BADA~4 performance models are used to reproduce the full trajectories.

\subsubsection{ATFM probabilities}
Similarly, the probability of having a regulation based on the time of the day, the ANSPs the flights are crossing, the capacity of the ANSPs, etc., is computed based on historical data as described in Section~\ref{subsec:ATFM_generator}. Different adjustments have been tested on the probabilities to adjust them with macroscopic quantities, such as delays.

\subsubsection{Passenger itineraries}
\label{sec:pax_calibration}
Possible itineraries for passengers are directly computed based on alliance structures and historical itineraries. However, the itinerary generator also needs some calibration regarding the number of connecting passengers and the load factors of the aircraft. This is done by generating new itineraries to ensure that load factors are within target windows, as explained in Section~\ref{sec:pax_generator}. The calibration is based on the data described in Section~\ref{subsec:calibration_input_data}.

Table~\ref{tab:pax_numbers} shows that, for the current scenario, there are over 3.4M passengers' itineraries modelled, of which around 8.4\% are connecting passengers (most of them with a single connection).


\begin{table}[htbp]
\begin{center}
\caption{Number of passengers modelled baseline scenario (current).}
\label{tab:pax_numbers}
\begin{tabular}{l|c|c|c}
&
Current &
L35 baseline &
L50 baseline\\
\hline
Total number passengers ('000) & 3 401 & 4 344 & 4 740\\
With 1 connection ('000) & 275 (8.1\%) & 373 (8.6\%) & 412 (8.7\%)\\
With 2 connections ('000) & 9 (0.3\%) & 12 (0.3\%) & 13 (0.3\%)\\
Total with connections ('000) & 284 (8.4\%) & 385 (8.9\%) & 425 (9.0\%)
\end{tabular}
\end{center}
\end{table}

\subsubsection{Flight schedules}
Calibration was also performed to assure realistic schedules for the additional flights created by the economic model. We first used a pattern analysis to study which kinds of sequences of airports the aircraft were following. We then predicted, for a given pattern, the initial time of departure of the first aircraft and all the subsequent times during the day. This was done by using a classification and regression tree (CART) trained on a month of data, with five predictors.

\section{Results}
\label{sec:results}

\correction{In this section we start by showing some results for the most important system-wide KPIs selected by the project. We then summarise other results that have been obtained with the model, in particular focusing on each of the three layers independently. Dedicated papers, with different focus areas regarding the results are also planned. }


\subsection{Key performance indicators}

\begin{figure*}[htbp]
\begin{center}
\includegraphics[width=\textwidth]{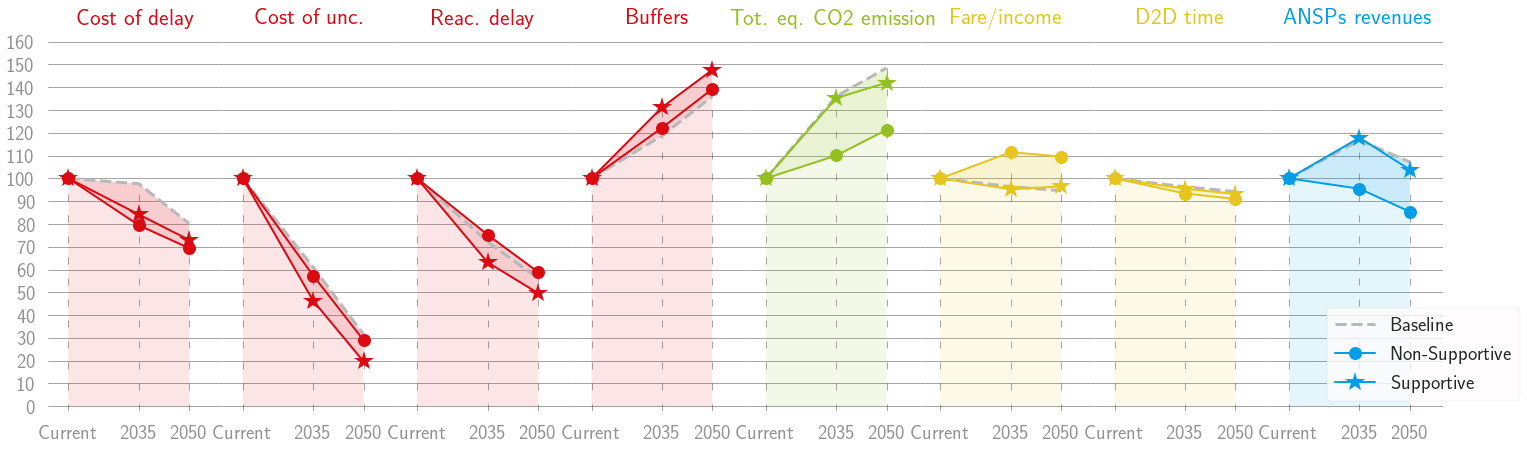}
\caption{Key indicators for stakeholders by scenario. From left to right: cost of delay per flight; cost of delay uncertainty per flight; reactionary delay per flight; buffer per flight; total equivalent CO$_2$ emissions; fare-to-income ratio; door-to-door travel time; ANSP revenues.}
\label{fig:nice_plot}
\end{center}
\end{figure*}

In Figure~\ref{fig:nice_plot} we show some key metrics for stakeholders and for the most complex scenarios, showing the baseline, `Supportive' and `Non-supportive' scenarios. 

\begin{correctionenv}Overall, in both scenarios, over the period we observe smaller costs of delay for the airlines, partly driven by smaller costs of uncertainty. This is mainly due to reactionary delays decreasing, this being in turn due to larger buffers. The situation is also quite positive for passengers: lower fare-to-income ratios, at least in the supportive scenario, and shorter door-to-door times. The situation is less positive for the environment. The model predicts a great increase in the total emissions in the future. This is mainly driven by the increase in traffic, but is also due to airlines using heavier aircraft and operating longer routes. Note that the model did not include any gains in efficiency due to better aircraft design (or switches to alternative power sources), so there is some scope for improvement in this regard. It is also interesting to track ANSPs' revenues, even though they are not an indicator \emph {per se}: the ANSPs are not in competition in the system and they are profit neutral.

\subsection{Other selected results}

\subsubsection{Impact of external parameters}
Deliverables \cite{del_51} and \cite{del_52} present further results, in which the impact of some parameters are explored more systematically. In particular, the price of fuel factor is analysed in detail. Very different values of the fuel price have been tested. The results indicated that airlines are greatly impacted by this price, but also that they transfer part of this extra cost to the passenger, which moderates the increase in traffic. This has a positive impact on the environment and is added to the benefits arising from flights using less fuel-intensive routes. 

\subsubsection{Observations by layer of the model}

We here briefly compare the results from the model layers and observe how they relate to each other.

For the strategic layer:
\begin{itemize}
    \item The strategic layer shows a tendency for airlines to increase the average size of their aircraft when capacity is scarce. This has already been observed in empirical data. 
    \item The operational cost decreases from 2035 to 2050, driven by various efficiency gains. In particular, uncertainty is halved, leading to a more stable and less costly environment for airlines. 
    \item Delays increase throughout this period, driven by the large increase in traffic, barely mitigated by the increase in airport capacity. ATFM capacity is also a problem, with a capacity crunch around 2035, mitigated by 2050 by efficiency improvements by ANSPs. 

\end{itemize}

For the pre-tactical layer:
\begin{itemize}
    \item The computation of flight plans indicates that the average flight plan distance tends to increase in the period from 2014 to 2050, leading to higher fuel burn per flight. However, the distance between origin and destination airports increases even more: longer routes are operated, but more efficiently. This leads to an increment in the average buffer time, as schedules have not been sufficiently tightened.
    \item The percentage of connecting passengers grows over time, from 8.1\% to 8.7\%. This might indicate that as the number of flights increases, the possibility of making connecting flights also does so, providing more alternatives to passengers on routes with insufficient demand to justify direct flights.
    \item Different operating costs might shift demand for particular airspace areas, resulting in variations in revenues per ANSP. The pre-tactical layer allows us to estimate such values. Most ANSPs obtain higher revenues than in the current scenario (linked with an increment in traffic), but these variations are not homogeneously experienced. 
\end{itemize}

In the tactical layer:
\begin{itemize}
    \item Delays decrease in 2035 and 2050. This is a out of kilter with the strategic layer, and arises from the fact that buffers are not properly taken into account by the strategic model. Buffer creation is a complicated issue, highly dependent on the operational environment and the airline business model. We plan to develop this line of research further in future.
    \item Average gate-to-gate times do not increase, despite the increment in delays. This is a volume effect, due to the fact that airlines tend, on average, to operate longer routes (in time) in 2035 and 2050.
\end{itemize}
\end{correctionenv}

\section{Conclusions and future work}
\label{sec:conclusions}

The Vista project has built a holistic model in order to capture high-level alignments and trade-offs between key indicators. The model is composed of three layers, aligned with the strategic, pre-tactical, and tactical ATM phases. 

The core of the strategic layer is the economic model, which allows us to capture complex feedback loops in the supply and demand interplay within the system. It is highly granular compared to standard economic models, using more than 45k interactive agents with their own objective functions and imperfect behaviours to build estimations of key features of potential futures. In particular, the model is able to capture hub versus point-to-point competition, without assuming prior archetypes for the agents. The pre-tactical layer generates individual flight plans, individual passenger itineraries and probabilities of experiencing ATFM regulations. These processes are based on the analysis of historical data (e.g., with route clustering and flight level preferences) to generate flight plans that are non-optimal, but closer to realistic, observed data. Passenger itineraries and ATFM regulations are also calibrated with historical data. The tactical layer uses these to execute a typical day of operations, including details at the passenger level in a door-to-door context. 

The model has been run on several scenarios, carefully built on various sources of data to represent various potential futures. In particular, scenarios including two degrees of support to the system have been chosen to highlight potential inconsistencies and interactions.

The results show that the model is able to describe a wide array of metrics at the same time. The strategic layer predicts moderately higher delays, a decrease in delay uncertainty, but overall increasing operational costs. This is partly due to airlines operating longer routes with heavier aircraft. Specific ANSPs' revenues and flight plan characteristics are captured by the pre-tactical layer including an increase in flight-plan buffers. The tactical layer shows that the increase in buffer times translates into smaller reactionary delays. This calls for a more detailed study of the interactions between the layers.

Vista was originally envisioned with an additional `learning loop' block, which would extract information from the tactical and pre-tactical layers and feed them back to the strategic layer. This would have allowed for a better alignment between the layers, even if one of the features of the model is indeed to have different levels of information between different blocks, as is the case in reality (when tactical execution needs to be thoroughly - but imperfectly - analysed to feed the strategic view of an airline, for example). This loop has not been built in Vista, but is an open possibility for the future of the model. In particular, it could use one of the multi-agents' learning paradigms which have recently been developed. The modular nature of the model also renders it open to future development and the flexible data input structure allows ready modification of the scenarios.

Another area of consideration is the sensitivity of the model to initial conditions and parameters. Whilst an analysis of the impact of various parameters has been performed, a more systematic approach is needed in order to fully explore the behaviours of the model(s), in particular the economic model.


As it is, the model can be used to produce high-level results on the potential trade-offs for the future of the ATM system. As highlighted, some system modifications tend to improve part of it and degrade others (e.g., with regard to the environment). Such quantitative analyses can be used to support evidence-led
policy making.

\section{Acknowledgements}
We thank the Deutsche Zentrum f\"{u}r Luft- und Raumfahrt (DLR) for kindly providing confidential formulae and valuable advice for the evaluation of emissions impacts, based on \cite{dahlmann}.

This project has received funding from the SESAR Joint Undertaking under grant agreement No 699390 under European Union's Horizon 2020 research and innovation programme. The opinions expressed herein reflect the authors' views only. Under no circumstances shall the SESAR Joint Undertaking be responsible for any use that may be made of the information contained herein.

\section*{Bibliography}
\bibliography{mybib}{}

\appendix
\renewcommand*{\thesection}{\Alph{section}}
\section{Annex -- ANSP capacity and delay}
\label{annex:ANSP}
In the model, we assume a relationship between average delay and average traffic within each ANSP. Contrary to the airports, we derive an explicit relationship based on a very simple queuing model.

First, we assume that the capacity $C$ of an ANSP is constant throughout the day for its airspace. Moreover, only the number of flights is taken into account towards its capacity (i.e. complexity is only a function of the traffic). Thus, the ANSP can have a maximum of $C$ flights in a given time window. Any extra flight is given a delay $\tau$ (the typical magnitude of an ATFM regulation). If $p$ is the p.d.f of the traffic through the day, then the average regulation delay will be:
\begin{equation}
\delta t = \frac{\tau}{C} \int_C^\infty (x-C)p(x)dx    
\end{equation}

For computational reasons, we wish to have an analytical relationship for this (in particular, because it needs to be inverted, see below). Hence, we assume that the traffic follows a normal distribution $\mathcal{N}(T, \sigma)$, which gives:
\begin{equation}
\delta t = \frac{\tau}{C} \left((T-C)F(T-C) + \sigma^2 p( T-C)\right),    
\end{equation}
with $F$ the corresponding cumulative distribution function. The variance $\sigma^2$ of the traffic distribution is also a function of the mean traffic within an ANSP. It is indeed expected that higher levels of traffic translate into higher levels of average delay and higher levels of uncertainty (variance). If one assumes that the main factor for the randomness of the traffic levels in the ANSPs' regions is the randomness of the departure delay, and using a simple Poisson model, it can be shown that: 
\begin{equation}
    \sigma = \sqrt{T} \left(1 - e^{-\frac{\sigma_{\delta t}}{\tau_0}}\right)\frac{1}{\tau}\left(1-\frac{1}{\tau}\right),
\end{equation}
where $\tau_0$ is the controlling time-window for the controller (from the definition of the declared capacity), set to 1 hour in the simulation, and $\sigma_{\delta t}$ is the standard deviation of the departure delay. 

Given the target delay, and having predicted a level of traffic (in the same way that the flights predict price), one needs to invert this equation in order to choose the capacity that is required to be below the delay target. This cannot be done analytically and is performed with a scalar minimisation algorithm using the Brent method. The result represents the ideal capacity that the ANSP should have to keep the delay under the target. This ideal capacity, however, cannot be always reached, because the ANSP has a maximum capacity. The maximum capacity represents the capacity that is achievable by the ANSP with a constant traffic structure and a given technology. Following insights from Belgocontrol (now Skeyes, in the consortium of the Vista project), it has been set to 120\% of the current capacity for every ANSP. This is a simplification, since in particular different ANSPs are already differently close to saturation. This maximum capacity depends only on the technology, and changes with various factors, in particular the SESAR-related ones.

Once its capacity has been fixed, the ideal capacity or its maximum one, whichever is smaller, the ANSP computes the corresponding cost $c$ and fixes the unit rate $u$, in order to have zero profit:
$$
u = 100 \frac{c}{V},
$$
where $V$ is the revenue, corresponding to $\sum_i d_i \sqrt{MTOW_i/50}$, with the sum computed over all flights crossing the ANSP’s region.

\section{Annex -- Airport delay/traffic relationship}
\label{annex:airport}
In the model, we use a simplified relationship between average delay at an airport and its average traffic:
$$
\bar{\delta t} = \delta t_0 + \frac{T}{C},
$$
where $\bar{\delta t}$ is the average delay `created' at the airport, $\delta t_0$ and $C$ are constants, with $C$ representing the capacity of the airport in a given time window, to be compared with the (mixed) traffic $T$ during this time. Both parameters are inferred from data, as explained below.

\begin{correctionenv}
The linear relationship between delay and traffic is a simple one, justified by queuing theory and our own statistical analysis.
Given a system, which can `process' an entity with a rate $\mu$ and under the assumption of a Poisson distribution for arrival entities in the system with a rate $\lambda$ (M/D/1 system type), the average waiting time (if $\lambda < \mu$) is given by $w = \lambda/(2\mu(\mu - \lambda))$. For airports, $\mu$ corresponds to their instantaneous capacity (maximum movements per unit of time), and $\lambda$ represents the traffic per unit of time. \correction{Usually, $\mu$ is much larger than $\lambda$ during the day, except during peak hours, and thus the equation can be approximated to $w = \lambda/(2\mu^2)$, where $2\mu^2$ plays the role of effective capacity in our model.

The assumption of Poisson-distributed arrivals is a strong one and its linearisation an even stronger one, but in practice we found that a linear relationship is acceptable for large airports, even when one includes peak hours}. In Figure \ref{fig:fits}, we show some examples of regressions for four airports. The inverse law arising from the Poisson hypothesis is sometimes able to capture high-delay phenomena but performs quite poorly in other cases. The exponential law, purely phenomenological, is more stable in this regard, but may overfit the data. The linear law is a good compromise, with less overfitting at high delay while fitting the rest of the data sufficiently well. Table \ref{tab:goodness_fit} shows metrics computed on the 20 largest airports in our dataset. The mean coefficient of determination is clearly much better for the linear law than for the others. Looking at the median, the exponential and inverse are only a little better. Hence, the linear law allows us to have a pretty good approximation with a stable fit over the airports, suitable for an automated procedure. \correction{This linear law works well in our case because we average the traffic over several days, smoothing out the divergence to some extent. Note also that this linear law is adopted largely due to its convenience: in practice, the model would work similarly with another law.}

\begin{figure}[htbp]
    \centering
    \includegraphics[width=\textwidth]{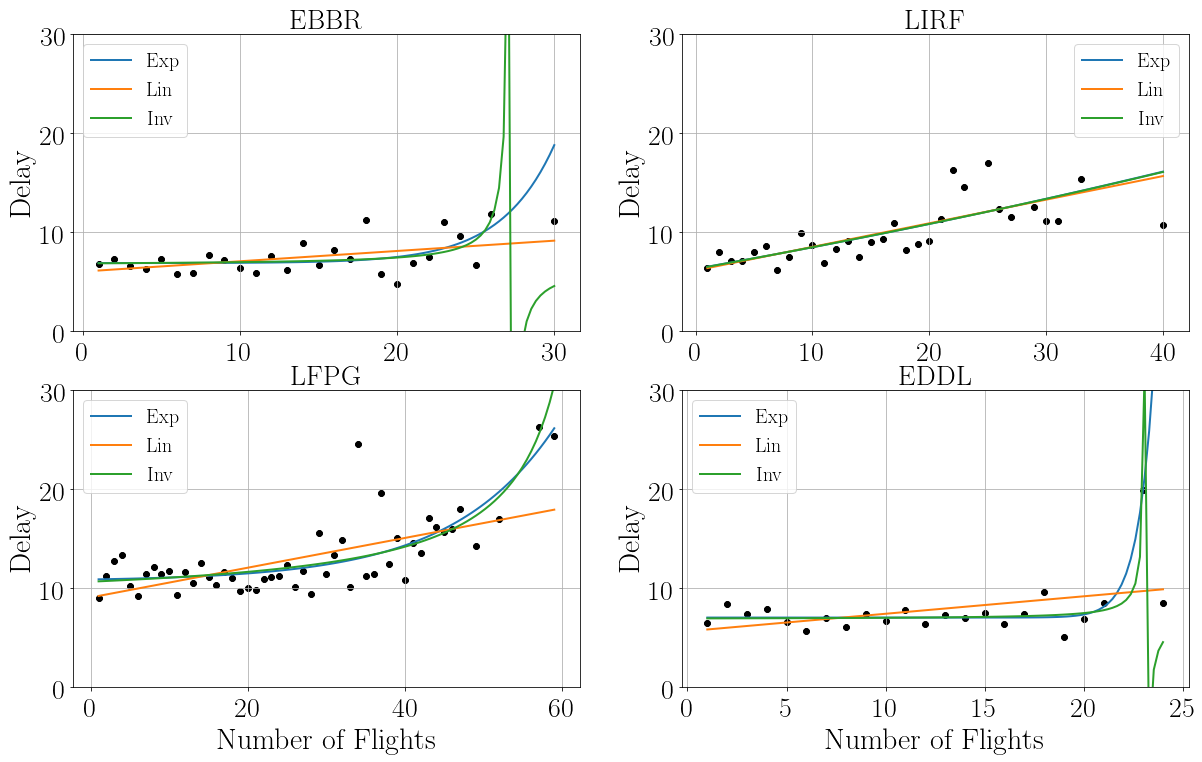}
    \caption{Three different regression laws (exponential, linear, and inverse, from the Poisson hypothesis), for four different airports. The black points represent hourly averages of traffic at the airport.}
    \label{fig:fits}
\end{figure}

\begin{table}[htbp]
\centering
\begin{tabular}{lrrrrrr}
\toprule
{} & \multicolumn{2}{l}{Exp} & \multicolumn{2}{l}{Inv} & \multicolumn{2}{l}{Lin} \\
{} &  $\chi^2$ &    $R^2$ &  $\chi^2$ &    $R^2$ &  $\chi^2$ &    $R^2$ \\
\midrule
Mean   &  1.43 &  0.19 &  1.44 & -0.94 &  1.44 &  0.35 \\
Median &  1.53 &  0.40 &  1.52 &  0.38 &  1.53 &  0.34 \\
\bottomrule
\end{tabular}
\caption{Mean and median of the adjusted $\chi^2$ and coefficient of determination $R^2$ for the different regression laws.}
\label{tab:goodness_fit}
\end{table}

To perform the regression and obtain the coefficients, we use CODA data, with a breakdown of the different causes of delay, and DDR data. In the economic model we have focused on the turnaround delay (CODA codes: 85, 86, 87, 97, 99), which can be thought as the delay generated only by airports (in particular, due to congestion).
\end{correctionenv}
However, in order to have different sets of coefficient per airport, we need more disaggregated data than just an average per airport, as in the CODA data. As a consequence, we used a proxy, namely the actual off-block time minus the initial off-block time, taken from the DDR data (AIRAC 1409). The mean delay from the DDR dataset is then shifted to correspond to the one in the CODA data.

To carry out the regression, we first compute the proxy of delay for a given time window, for instance one minute, for the whole AIRAC. We compute the number of departures during the same time window and we perform a linear regression. We compute the goodness of fit (using $R^2$ as score), and we then sweep the width of the time window and perform the regression for each of them. The width of the time window which best fits the data is selected, and the corresponding parameters selected for this airport. Interestingly, the typical optimal time window is between 30 and 90 minutes, which can be regarded as the typical response time of the airport from a dynamical point of view. After this step, many airports have quite a high goodness of fit, but some are still poorly fitted. We decided to set the parameters for them to the mean values of the best fits among the other airports. There might be a bias in so doing, since the best fits are usually given by the busiest airports.


\end{document}